\documentclass[twocolumn,showpacs,aps,prl,superscriptaddress]{revtex4} %\documentclass[11pt]{article}
\usepackage{graphicx}
\usepackage{dcolumn}
\usepackage{amsmath}
\usepackage{epsfig}

\newcommand{\BABARPubYear}    {06}
\newcommand{\BABARPubNumber}  {068}

\newcommand{\SLACPubNumber} {12335}

\input pubboard/babarsym.tex

\def\Bztokspp{\ensuremath{B^0\to\KS\pi^0\pi^0}\xspace}
\def\scp{\ensuremath{S}\xspace}
\def\ccp{\ensuremath{C}\xspace}
\def\sf{\ensuremath{S}\xspace}
\def\cf{\ensuremath{C}\xspace}

\def\finalscp{\ensuremath{0.72\pm 0.71\,\pm 0.08} \xspace}
\def\finalccp{\ensuremath{0.23\pm 0.52\ \pm 0.13} \xspace}

\def\Btag{\ensuremath{B_{\rm tag}}\xspace}

\def\Bcp{\ensuremath{B_{CP}}\xspace}

\def\figurebox#1#2#3{%
    \def\arg{#3}%
    \ifx\arg\empty
    {\hfill\vbox{\hsize#2\hrule\hbox to #2{\vrule\hfill\vbox to #1{\hsize#2\vfill}\vrule}\hrule}\hfill}%
    \else
    {\hfill\epsfbox{#3}\hfill}%
    \fi}

\begin{document}

\noindent

\begin{flushleft}

\preprint{\babar-PUB-\BABARPubYear/\BABARPubNumber} 
\preprint{SLAC-PUB-\SLACPubNumber} 
{\bf \babar-PUB-\BABARPubYear/\BABARPubNumber} 
{\bf SLAC-PUB-\SLACPubNumber} 
\end{flushleft}

\title{
{\large \bf Measurement of \boldmath{\CP} Asymmetry in 
  \boldmath{$B^0\to \KS \pi^0\pi^0$} Decays}
}

%% author list as of 01-Nov-2006 (581 authors)
%
\author{B.~Aubert}
\author{M.~Bona}
\author{D.~Boutigny}
\author{Y.~Karyotakis}
\author{J.~P.~Lees}
\author{V.~Poireau}
\author{X.~Prudent}
\author{V.~Tisserand}
\author{A.~Zghiche}
\affiliation{Laboratoire de Physique des Particules, IN2P3/CNRS et Universit\'e de Savoie, F-74941 Annecy-Le-Vieux, France }
\author{E.~Grauges}
\affiliation{Universitat de Barcelona, Facultat de Fisica, Departament ECM, E-08028 Barcelona, Spain }
% ...
\collaboration{The \babar\ Collaboration}
\noaffiliation

\date{\today}

\begin{abstract}
We present a measurement of the time-dependent \CP 
asymmetry for the neutral $B$-meson decay into the $CP = +1$ 
final state $\KS\pi^0\pi^0$,
with $\KS\to \pi^+\pi^-$. 
We use a sample of approximately 227 million $B$-meson pairs recorded 
at the $\Upsilon(4S)$ resonance with the \babar\ detector at the \pep2\ 
$B$-Factory at SLAC. From an unbinned maximum
likelihood fit we extract the mixing-induced \CP-violation
parameter $\scp = \finalscp$ and the direct \CP-violation
parameter $\ccp = \finalccp$, where the first uncertainty
is statistical and the second systematic.
\end{abstract}


\maketitle

\CP violation effects in decays of $B$ mesons dominated by
$\b\to\s\qbar\q$ transitions ($\q=\u,\d,\s$), are
potentially sensitive to contributions from physics beyond the
standard model (SM)~\cite{ref:newphysics}. The \B{}-factory
experiments have explored time-dependent \CP{}-violating (CPV)
asymmetries, occuring due to a phase difference between
mixing and decay amplitudes, in several such decays~\cite{cc}, including $B^0\to\phi
K^0${}~\cite{Abe:2003yt,Aubert:2004ii},
$B^0\to\KS\KS\KS$~\cite{ksksks}, $B^0\to\eta^\prime
\KS${}~\cite{Abe:2003yt,Aubert:2003bq}, $B^0\to K^+ K^-
\KS${}~\cite{Abe:2003yt,Aubert:2004ta}, $\Bz\to
f_{0}(980)\KS$~\cite{Aubert:2004am} and
$B^0\to\KS\piz$~\cite{Aubert:2004xf}.  Within the SM, the 
magnitude of the CPV asymmetry in
these decays is expected to be approximately equal to the one in
$\b\to\c\cbar\s$ decays, such as $\Bz \to \jpsi
\KS$~\cite{ref:newphysics}.
A major goal of the $B$-factory experiments
is to reduce the experimental uncertainties of these measurements and
to add more decay modes in order to improve the sensitivity to
beyond-the-SM effects.

In this paper we present a measurement of the CPV asymmetry
in the decay $B^0\to\KS\piz\piz$. The $\KS\pi^0\pi^0$ final state is
a \CP -even eigenstate, re\-gard\-less of any resonant
substructure~\cite{Gershon:2004tk}. 
In the SM this decay is dominated by the $b\to \s\qbar\q$ weak amplitude,
with $q=u,d$, and we expect $\scp\simeq
-\sin 2\beta$ and $\ccp \simeq 0$~\cite{ref:newphysics}.  Here \ccp
and \scp are respectively the 
magnitudes of CP violation in the decay and in the interference 
between decay and mixing, and the angle $\beta$ is
defined as $\beta$ = $arg(- V_{cd} 
V^*_{cb}/V_{td}V^*_{tb})$, where $V_{ij}$ are the elements of the
Cabibbo-Kobayashi-Maskawa (CKM) quark-mixing 
matrix~\cite{ckm}.
A possible contribution from a tree-level
$b\to u\bar{u}s$ amplitude is doubly Cabibbo suppressed with respect
to the leading gluonic penguin diagram.

The data used in this analysis were collected with the 
\babar\ detector~\cite{ref:babar} at the PEP-II asymmetric-energy
$e^+e^-$ collider ~\cite{pep}.  A sample of
$226.6\pm2.5$ million \BB\ pairs was recorded at the $\Upsilon (4S)$
resonance center-of-mass energy $\sqrt{s}=10.58\ \gev$. 
 The \babar\ detector is described in detail
elsewhere~\cite{ref:babar}. Charged particles are detected 
and their momenta measured by the combination of a silicon 
vertex tracker (SVT), consisting of five layers
of double-sided detectors, and a 40-layer central drift chamber,
both operating in a 1.5 T solenoidal magnetic field.
Charged-particle identification is provided by measurements of energy
loss in the tracking devices and by an internally reflecting ring-imaging
Cherenkov detector covering the central region.
Photons and electrons are detected by an elec\-tro\-mag\-ne\-tic 
calorimeter (EMC) composed of 6580 CsI(Tl) crystals.  The typical resolution 
for the $\piz$ signal in the $\gamma\gamma$ invariant mass spectrum is 
better than 7~\mevcc.

We search for \Bztokspp decays in neutral $B$ meson candidates selected
 using charged-particle multiplicity and event
topology~\cite{ref:Sin2betaPRD}.  We reconstruct $\KS\to\pip\pim$
candidates from pairs of oppositely charged tracks.  The two-track
combinations must form a vertex with a $\chi^2$ probability greater
than $0.001$ and a $\pip\pim$ invariant mass within
11.2~\mevcc of the nominal \KS\ 
mass~\cite{Eidelman:pdg2004}.  We form $\piz\to\gamma\gamma$
candidates from pairs of photon candidates in the EMC, where each photon
is isolated from any charged track, carries a minimum energy of
30~\mev, and has the expected lateral shower shape.
B meson candidates are formed from $\KS\piz\piz$ combinations
and constrained to originate from the \epem{} interaction region using
a geometric fit. We require that the $\chi^2$ probability of the 
fit, which has one degree of freedom, is greater than $0.001$.  We
extract the \KS{} decay length $L_{\KS}$ and the $\piz\to\gamma\gamma$
invariant mass from this fit and require 110 $< m_{\gamma\gamma}
<$ 160~\mevcc and $L_{\KS}$ greater than five times its uncertainty.
The cosine of the angle between the direction of the decay photons 
in the center-of-mass system of the mother \piz and the \piz  flight
direction in the lab frame must be less than 0.92.

We reconstruct a \Bz\  decaying into the \CP
eigenstate $\KS \pi^0\pi^0 $ ($B_{CP}$) and the vertex and flavor of
the other $B$ meson ($B_{\rm tag}$).
The difference $\deltat \equiv t_{\CP} - t_{\rm tag}$
of the proper decay times is obtained from the measured distance between the $B_{\CP}$
and  $B_{\rm tag}$ decay vertices and from the boost ($\beta \gamma =0.56$) of 
the \epem system. Ignoring resolution effects, the \deltat\ distribution is given by:
\begin{eqnarray}
\label{eq:cpt}
  {\cal P}_\pm(\Delta t) &=& 
        \frac{e^{-\left|\deltat\right|/\tau_{\Bz}}}{4\tau_{\Bz}} [1
        \mp\Delta w \pm \label{eq:FCPdef} \\
   &&\hspace{-1em}(1-2w)\left( \sf\sin(\deltamd\deltat) -
\cf\cos(\deltamd\deltat)\right)]. \nonumber
\end{eqnarray}

The upper (lower) sign denotes a decay accompanied by a \Bz (\Bzb) tag,
$\tau_{\Bz}$ is the mean neutral $B$ lifetime, $\deltamd$ is the mixing
frequency, and the mistag parameters $w$ and
$\Delta w$ are the average and difference, respectively, of the probabilities
that a true $\Bz$\ is incorrectly tagged as a $\Bzb$\ and vice versa.
The tagging algorithm \cite{s2b} has seven mutually exclusive tagging 
categories of
differing purities including one for untagged events that we
retain only for yield determinations.  The effective tagging efficiency, defined as 
the tagging efficiency times $(1-2w)^2$ summed over all categories, is
$( 30.5\pm 0.6)\%$, 
as determined from a large sample 
of $B^0$-decays to fully-reconstructed flavor eigenstates ($B_{\rm flav}$).

We use the same technique
developed for $B^0\to\KS\piz$ decays of Ref.~\cite{Aubert:2004xf}
to reconstruct the \Bztokspp vertex using the
knowledge of the \KS trajectory and the average 
interaction point (IP) in a geometric fit.  The extraction of \deltat has been
extensively validated in data~\cite{Aubert:2004xf}, and on large
samples of simulated \Bztokspp decays with different values of \sf and
\cf. 

The per-event estimate of the uncertainty on \deltat{}, $\sigma(\Delta t)$, reflects the
strong dependence of the \deltat resolution on the $\KS$ flight
direction and on the number of SVT layers traversed by the $\KS$ decay
daughters. In about 70\% of the events both pion tracks 
are reconstructed from at least 4 SVT hits, leading to sufficient resolution
for the time-dependent measurement. The average \deltat{} resolution,
$\sigma_{\deltat}$, 
in these events is about 1.0~ps. For events that have fewer than 4 SVT
hits or for which $\sigma(\Delta t) > 2.5$~ps or
$\deltat>20$~ps, the \deltat information is not used.
However, since \cf can also be extracted from flavor tagging
information alone, these events still contribute to the measurement 
of \cf.

We extract the signal yield, \sf and \cf
from an unbinned extended maximum likelihood 
fit where we parameterize the distributions of several
kinematic and topological variables for signal and background events
in terms of probability density functions (PDFs).

For each $B$ meson candidate we compute two kinematic variables, 
the energy difference $\Delta E = E_B^* - \frac{1}{2}\sqrt{s}$ and
the beam-energy--substituted mass 
$\mes = \sqrt{(\frac{1}{2}s + \vec{p}_0\cdot\vec{p}_B)^2/E_0^2 - p_B^2}$
\cite{ref:babar}, where the 
subscripts 0 and $B$ refer to the initial 
$\Upsilon(4S)$ and the $B_{CP}$ candidate in the lab frame,
respectively, and the  
asterisk denotes the \epem center-of-mass frame.
For signal events, \DeltaE is expected to peak at zero and 
\mes at the known $B$ meson mass. From a detailed simulation  we
expect  a signal resolution of about 3.6~\mevcc in \mes and 45~MeV in
$\Delta E$. Both distributions exhibit a 
low-side tail due to the response of the EMC to
photons. We remove a small dependence 
of the signal $\Delta E$ resolution on the location in the
$\KS\piz\piz$ Dalitz plot
by using $\Delta E/\sigma(\Delta E)$ as a discriminating variable
instead of $\Delta E$, where 
$\sigma(\Delta E)$ is the calculated uncertainty in $\Delta E$. We select
candidates with $\mes > 5.20$~\gevcc and $-5 < \Delta E/\sigma(\Delta E) < 2$.
To suppress other $B$ meson decays we also require $-0.25 < \Delta E < 0.1$~GeV,
which does not affect the signal $\Delta E/\sigma(\Delta E)$ distribution.

The background $B$ meson candidates come primarily from random combinations 
of \KS and neutral pions produced in continuum events of the type
$e^+e^-\to q\bar{q}$, where $q = u,d,s,c$. 
Background from  $\BB$ events may occur either in charmless decays
with a \KS as a decay product, or from decays where the \KS is from an
intermediate charmed particle.  
The shapes of event variable distributions are obtained from signal 
and background Monte Carlo (MC) samples and high statistics data control samples.
The charmless $B$ background forms a broad peak in \mes near the $B$-meson  
mass; other $B$ background distributions do not peak in \mes.  None of the
$B$ backgrounds peak in the $\Delta E/\sigma(\Delta E)$ distribution.

We reduce continuum background events, while 
retaining 90\% of the signal, by requiring  
$|\cos\theta_T| < 0.9$, where $\theta_T$ is the angle between the 
thrust axis of the $B_{CP}$ candidate's decay products and the thrust 
axis formed from the other particles in the event.  We combine 
$\theta_T$, the angle between the $B_{CP}$ momentum 
and the beam axis, $\theta_B$, and the sum of the momenta $\vec{p_i}$ of the other
particles in the event weighted by the Legendre polynomials 
$L_0(cos(\theta_i))$ and $L_2(cos(\theta_i))$ in a neural network (NN).  
The $NN$ has two hidden layers with 4 neurons each, and is trained 
and evaluated \cite{bfgs} on different subsets of simulated signal and 
continuum events and on data taken about 40~\mev below the nominal 
center-of-mass energy. To parameterize the $NN$ shape, we divide the $NN$ 
output into intervals, chosen such that  
they are uniformly populated by signal events (see, e.g.,
Ref. ~\cite{btopipi}).

We suppress background from other $B$ decays by excluding several invariant 
mass  intervals: $m(\KS\piz) > 4.8$~\gevcc eliminates $B^0\to\KS\piz$, 
$1.75 <  m(\KS\piz) < 1.99$~\gevcc reduces $B^0\to\bar{D}^0\piz$ 
to fewer than 10 expected candidates, $m(\piz\piz) < 0.6$~\gevcc removes 
$\eta\KS$ and $\eta^\prime \KS$, and $3.2 < m(\piz\piz) < 3.5$~\gevcc removes
$\chi_{c0}\KS$ and $\chi_{c2} \KS$ candidates.

The 
signal reconstruction efficiency after all of the above requirements
is about 15\%. Based on MC simulations 
we expect more than one $B_{CP}$ candidate in 13\% 
of the signal events. The selection of the best candidate is 
based only on \piz information, since the number of multiple \KS
candidates is negligible (less than 0.1\%).  We select the candidate
whose two $\piz$s have masses that are closest to the expected value.  

For each selected \Bztokspp candidate we examine the remaining tracks 
in the event to determine the decay vertex position~\cite{ref:Sin2betaPRD}  
and the flavor of \Btag \cite{s2b}. 
We parameterize the performance of the tagging 
algorithm in a data sample of fully reconstructed 
$\Bz\to D^{(*)-}\pip/\rho^+/a_1^+$ decays.
For the continuum background, the fraction of events in each tagging
category is extracted from a fit to the data. 

By exploiting regions in data that are dominated by background,
and by using simulated events, we verify that 
the observables are sufficiently independent that 
we can construct the likelihood from the product of one-dimensional 
PDFs, apart from the signal \mes and $\Delta E/\sigma(\Delta E)$
which are correlated.  For these observables, we use a two-dimensional 
PDF derived from a smoothed, simulated distribution.
We obtain the PDF for the \deltat of signal events from the
convolution of Eq.(\ref{eq:cpt}) with a resolution function 
${\cal R}(\delta t$ $\equiv \deltat -\deltat_{\rm true},$
$\sigma_{\deltat})$, where $\deltat_{\rm true}$ is the actual
$\deltat$ in the simulated event.
The resolution function is parameterized as the sum of two Gaussians
with a width proportional to the reconstructed $\sigma_{\deltat}$, and
a third Gaussian with a fixed width of
8~ps~\cite{ref:Sin2betaPRD}. The first two Gaussian distributions have a
non-zero mean, proportional to $\sigma_{\deltat}$, to account for a bias induced
by charm decays on the \Btag side.  We
have verified in simulations that the parameters of ${\cal R}(\delta t,
\sigma_{\deltat})$ for \Bztokspp events are similar to those
obtained from the $B_{\rm flav}$ sample, even though the distributions
of $\sigma_{\deltat}$ differ considerably. We therefore extract these
parameters from a fit to the $B_{\rm flav}$ sample. 
We also use this resolution function for the description of background
from other charmless $B$ decays.
While the resolution functions for $B$ decays into
open charm final states and continuum have the same
functional form as used for signal events, the parameters for the
\deltat PDF of the open-charm background are determined from MC
simulation and they are varied in the fit to data for the continuum.

We subdivide the data into the tagging categories $k$, events with and 
without \deltat information (sets I and II), and those events
located in the inside or  
outside region of the Dalitz plot ($in$ and $out$). The last
subdivision accounts for the 
higher contribution and different characteristics of continuum
background near the Dalitz  
plot boundary. We define the quantity $\delta = min(m_{12}^2,m_{13}^2,m_{23}^2)$,
where $m_{ij}$ is the invariant mass of the $B_{CP}$ decay daughters $i$ and $j$
combined. This $\delta$ corresponds to the distance of an event in the
Dalitz plot to the nearest 
Dalitz plot boundary in the limit of massless daughters. We split the data 
at $\delta = 3.5$~GeV$^2/$c$^4$. 

We maximize the logarithm of the extended likelihood
$\mathcal{L} = e^{-(N_S + N_B)} \cdot \prod_{k=1}^{7} \calL_k$
with $N_S$ and $N_B$ $= \sum_B n_B$ the total signal and background yields, 
respectively. The likelihood $\calL_k$ in each tagging category $k$
(with tagging fraction $\epsilon_k$) is given as: 
\begin{align*}
{\cal L}_k =
\prod_{j}^{N\mbox{\tiny I out k}} \Bigl[ &
N_S\, \epsilon^S_k f^S_I f^S_{out}\, P^S_{k,j} +\\
&\sum_B n_B\, \epsilon^B_k f^B_I f^B_{out}\, P^B_{k,out,j}  \Bigr] \times \\
\prod_{j}^{N\mbox{\tiny I in k}} \Bigl[ & 
N_S\, \epsilon^S_k f^S_I (1 - f^S_{out})\, P^S_{k,j} + \\
  & \sum_B n_B\, \epsilon^B_k f^B_I (1 - f^B_{out})\, P^B_{k,in,j}  \Bigr] \times \\
\prod_{j}^{N\mbox{\tiny II out k}} \Bigl[ &
N_S\, \epsilon^S_k (1 - f^S_I) f^S_{out}\, Q^S_{k,j} + \\
  & \sum_B n_B\, \epsilon^B_k (1 - f^B_I) f^B_{out}\, Q^B_{k,out,j}  \Bigr]  \times \\
\prod_{j}^{N\mbox{\tiny II in k}} \Bigl[ &
N_S\, \epsilon^S_k (1 - f^S_I) (1 - f^S_{out})\, Q^S_{k,j} + \\
 & \sum_B n_B\, \epsilon^B_k (1 - f^B_I) (1 - f^B_{out})\, Q^B_{k,in,j}  \Bigr] .
\end{align*} 
The probabilities $P^S$ ($Q^S$) and $P^B$ ($Q^B$) for each measurement $j$ are the 
products of PDFs for signal ($S$) and background ($B$) classes:
  $P_{k} = PDF(m_{ES}, \Delta E/\sigma(\Delta E))\cdot 
  PDF(NN)\cdot PDF(\Delta t,\sigma(\Delta t), \mbox{tag}_{k}, k)$,
where for the background  $PDF(m_{ES}, \Delta E/\sigma(\Delta E))$
= $PDF(m_{ES})\cdot PDF(\Delta E/\sigma(\Delta E))$.
The probabilities $Q$ do not depend on $\Delta t$ and $\sigma (\Delta t)$
and are used to extract \cf from the yields.
The fractions of events with \deltat information for signal and
background, $f^S_I$ and $f^B_I$, and fractions of events 
in the outside Dalitz plot region, $f^S_{out}$ and $f^B_{out}$, are
varied in the fit except for the fractions for B backgrounds which
are determined from simulation. 
For about 22\% of our signal $B$ candidates one or two of the \piz
decay photons associated with \Bcp originate from the \Btag . According to Monte Carlo 
simulation studies we expect to measure the same 
\sf and \cf  in these cross-feed events as in the correctly
reconstructed signal ($true$) since the contribution 
of the \piz to the \deltat measurement is marginal. 
To account for differences in the PDF distributions for the signal
probabilities $P^S$ ($Q^S$) we define the signal probability to be a
linear combination of the correctly reconstructed signal and
cross-feed events with the relative weight determined from simulation.
Parameters of signal PDFs are the same for the different Dalitz plot regions.
The PDFs for $B$ backgrounds are identical for the Dalitz inside and
outside regions.
The tagging fractions for the signal and the $B$ decay backgrounds 
are the same, while those of the continuum background are different.

\begin{figure}[!htb]
\begin{center}
\begin{tabular}{cc}
\includegraphics[height=3.0cm]{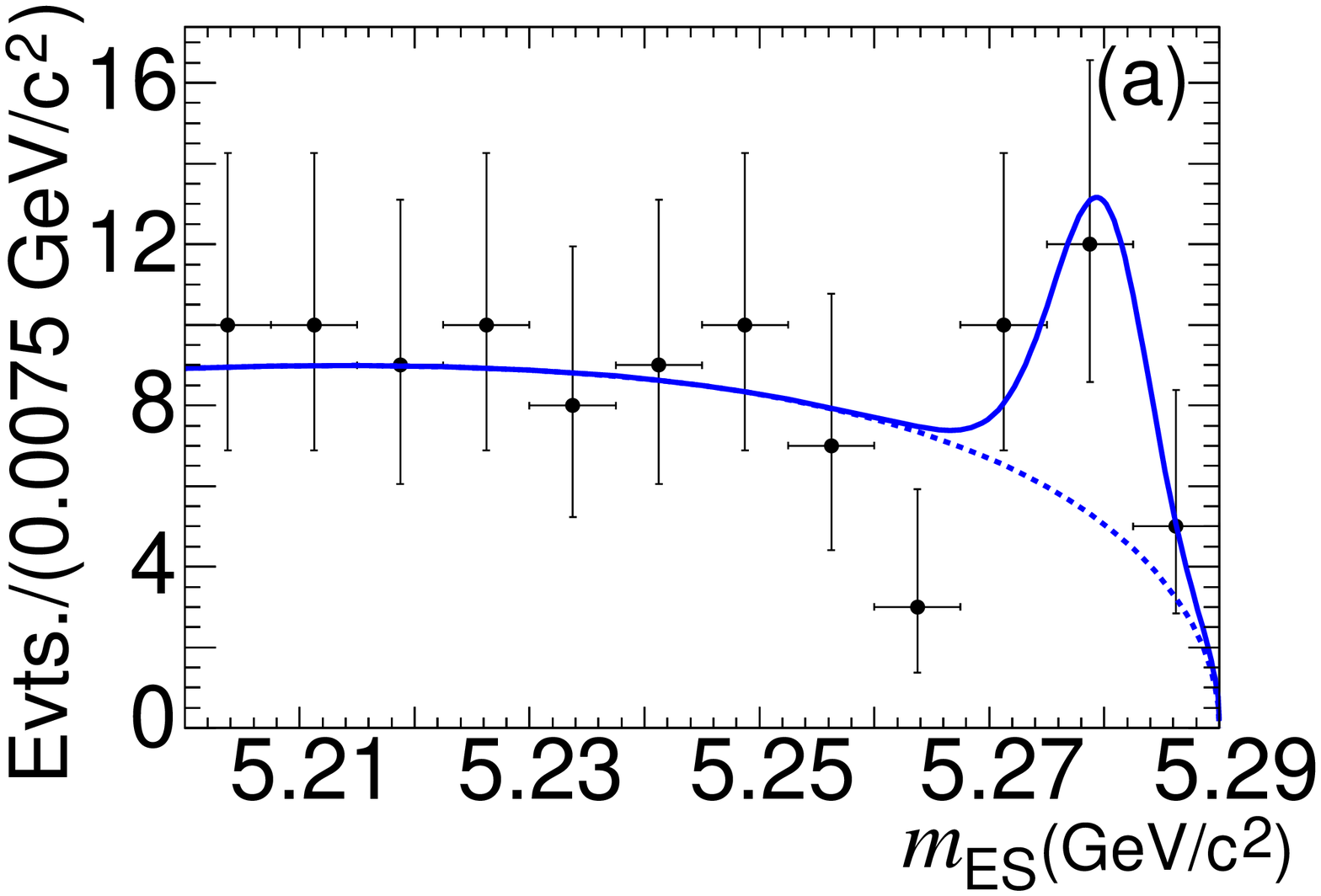} &
\includegraphics[height=3.0cm]{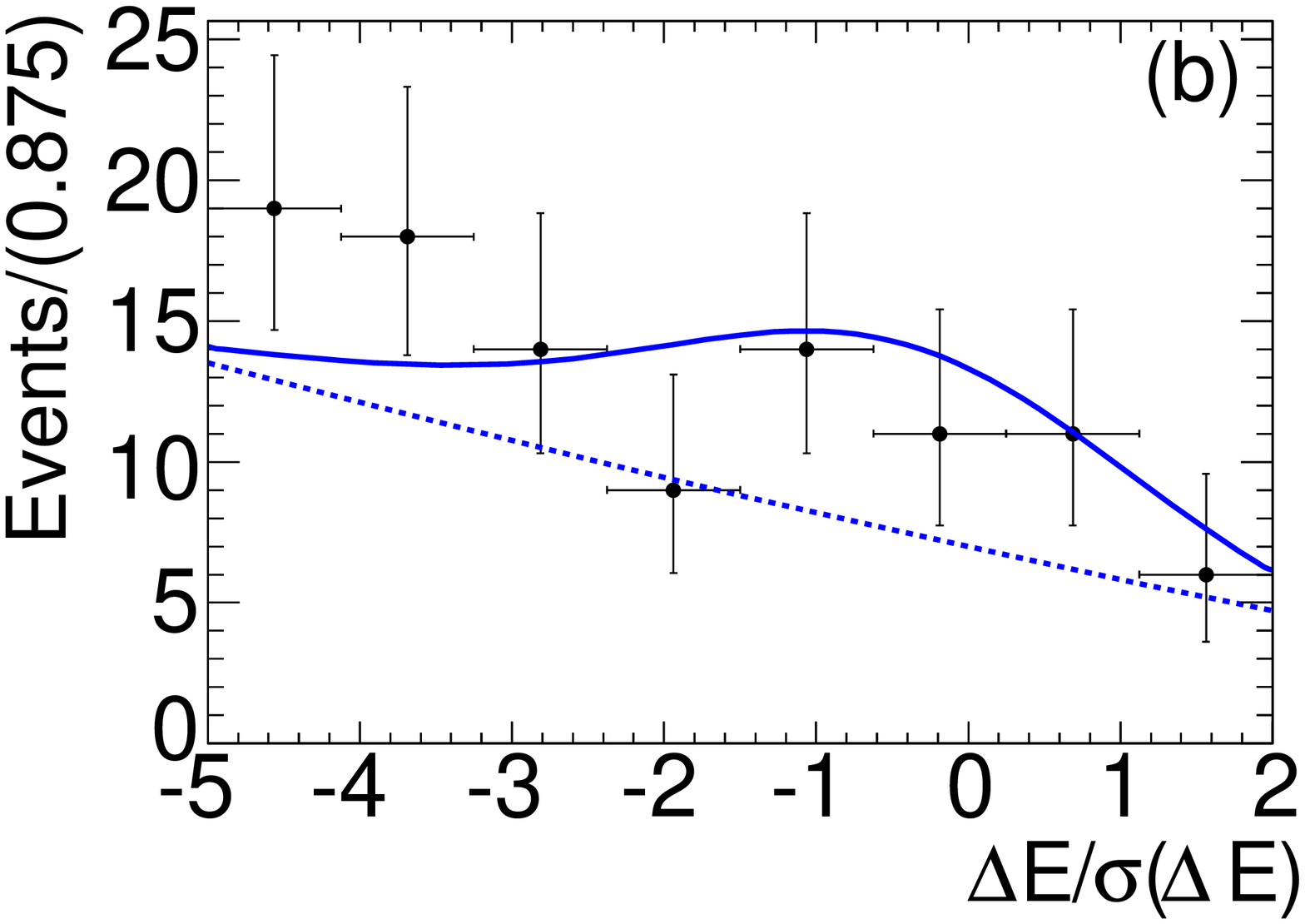} \\
\includegraphics[height=3.0cm]{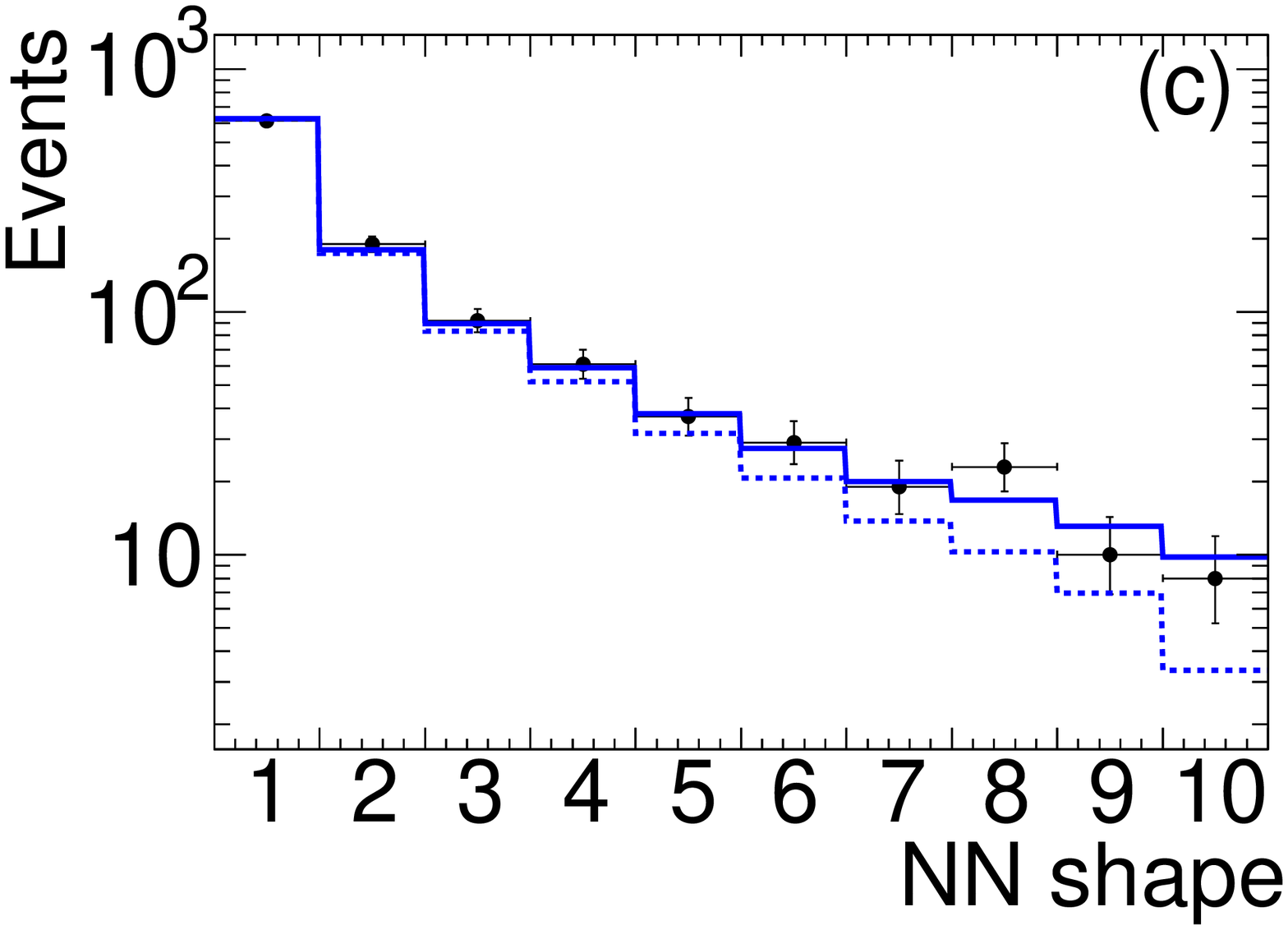} & 
\includegraphics[height=3.0cm]{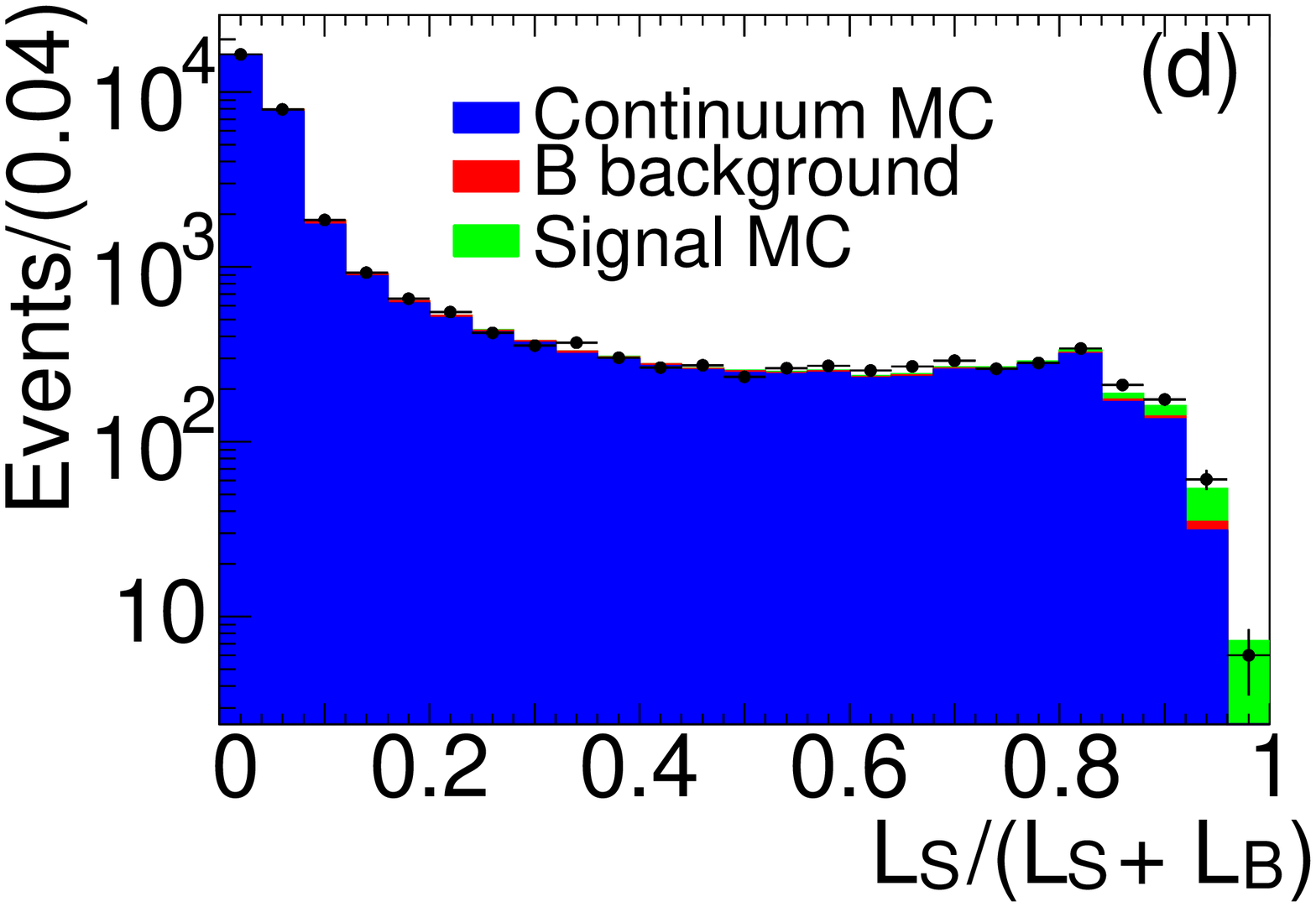} \\
\includegraphics[height=3.0cm]{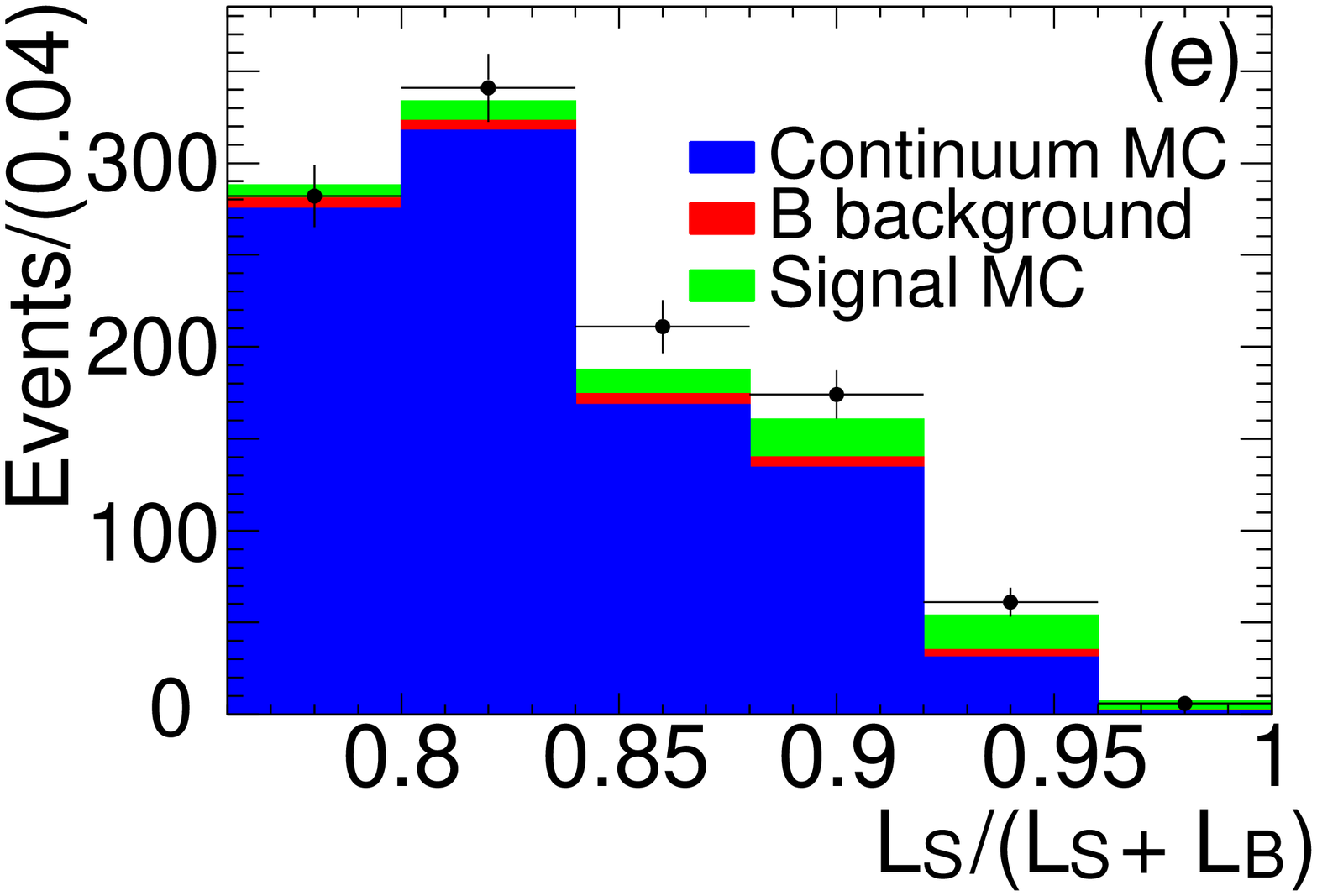} & \\
\end{tabular}
\caption{Distribution of the event variables (a) \mes, (b) $\Delta E/\sigma(\Delta E)$, 
and (c) $NN$ output in 10 bins after reconstruction and a requirement 
on the ratio of signal likelihood to the signal-plus-background likelihood, calculated 
without the plotted variable. The solid line represents the fit result
for the total event yield and the dotted line for the total background.  
Plot (d) shows the ratio of the signal likelihood to signal-plus-background
likelihood with all variables included, data (dots) with the fit 
result superimposed. Plot (e) shows the same quantity as (d) 
close to one and with a linear scale. 
\label{fig1}
}
\end{center}
\end{figure}

\begin{figure}[!htb]
\begin{center}
\includegraphics[height=7.7cm]{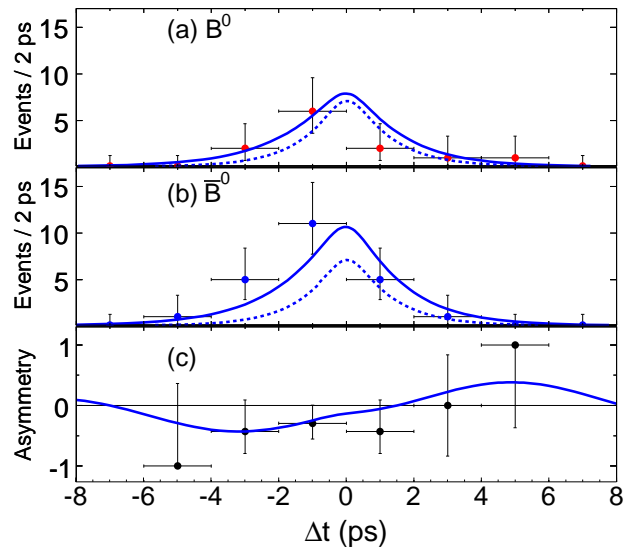} 
\caption{
Plots (a) and (b) show the \deltat distributions of $B^0$- and
$\bar{B}^0$-tagged \Bztokspp candidates. The solid lines refer
to the fit for all events; the dashed lines correspond to the 
total background. Plot (c) shows the raw asymmetry (see text).
A requirement is applied on the event likelihood to suppress 
background.
\label{fig2}
}
\end{center}
\end{figure}

The central values of \sf and \cf were hidden until the analysis was complete.
From a data sample of 33 058 \Bztokspp candidates, we find $N_S = 117\pm 27$
signal decays with \scp = \finalscp and \ccp = \finalccp where the first uncertainty
is statistical and the second systematic. 
The linear correlation coefficient between the two \CP parameters is
2\%, and the statistical significance of the signal yield is $5.8\sigma$.
The yield of charmless $B$ background is consistent with zero, and
the fraction of the signal in the outside Dalitz region is $0.78 \pm 0.07$.
Figure~\ref{fig1} shows the distributions of the event variables
\mes, $\Delta E/\sigma(\Delta E)$, and $NN$ output, and the ratio of the
signal likelihood to signal-plus-background 
likelihood with all variables included.
Figure~\ref{fig2} shows the \deltat distributions for
the $B^0$- and the $\bar{B}^0$-tagged subsets, and the raw asymmetry 
$[N_{B^0} - N_{\bar{B}^0}]/[N_{B^0} + N_{\bar{B}^0}]$, where
the $N_{B^0}$ ($N_{\bar{B}^0}$) is the number of $B^0$
($\bar{B}^0$) -tagged events.
In all plots, data are displayed together with the result from the fit after applying 
a requirement on the ratio of signal likelihood to signal-plus-background likelihood 
(computed without the variable plotted) to reduce the background. 

We consider the systematic uncertainties listed in Table~\ref{syst}.
These include the uncertainties in the parameterization of PDFs for
signal and backgrounds which were evaluated by varying parameters
within one standard deviation or using alternative shape functions.
The largest contribution to the uncertainty for \cf is caused by the
$NN$ shape for 
continuum inside the Dalitz plot and
for \sf from the 2-D parameterization of \mes and $\Delta
E/\sigma(\Delta E)$.
We consider uncertainties in the background fractions and \CP 
asymmetry in the charmless $B$ background,
the parameterization of the \deltat resolution function and
the vertex finding method, knowledge of the event-by-event beam spot
position, imprecision in the SVT alignment, and the possible interference
between the suppressed $\bbar\to\ubar\c\dbar$ amplitude with the favored
$\b\to\c\ubar\d$ amplitude for tag-side $B$-decays~\cite{kirkby}.
We fix $\tau_{\Bz} = 1.532$~ps and $\Delta m_d = 0.505$~ps$^{-1}$
and vary them by one standard deviation~\cite{Eidelman:pdg2004}.
We correct for the small fit bias which is determined 
using fits to a large number of simulated experiments, where
signal and backgrounds are mixed together in the expected proportions.
The uncertainty of the fit bias is accounted for as a systematic error.

We perform several consistency checks, including the measurement of the
\Bz lifetime; we obtain $\tau_{\Bz} = 1.25\pm 0.47$~ps. We embed different
$B$ background samples from Monte-Carlo simulation in the data sample
and obtain consistent yields and \CP parameters from the fit.

\begin{table}[!b]
\caption{Sources of systematic uncertainties on \sf and \cf. 
The total error is obtained by summing the individual 
errors in quadrature.}
\begin{center}
\begin{tabular}{|l|c|c|}
\hline
Source & $\sigma (\sf)$ & $\sigma (\cf)$ \\
\hline
Signal and background PDF parameterization      & 0.05  & 0.11 \\
Background fractions                            & 0.03  & 0.02 \\
\CP in charmless $B$ background                 & 0.03  & 0.01 \\
Vertex finding/resolution function              & 0.02  & 0.05 \\
Beam spot position                              & 0.00  & 0.00 \\
SVT alignment                                   & 0.02  & 0.01 \\
Tag side interference                           & 0.00  & 0.01 \\
$\Delta m_d$, $\tau_B$                          & 0.02  & 0.01 \\
Fit Bias                                        & 0.04  & 0.02 \\
\hline
Total systematic error                          & 0.08  & 0.13 \\
\hline
\hline
\end{tabular}
\end{center}
\label{syst}
\end{table}

In summary, we
measure the \CP violating 
asymmetries in \Bztokspp ($\KS\to\pi^+\pi^-$) decays
reconstructed from a sample of approximately 227 million $B\bar{B}$ pairs. 
From an unbinned extended maximum likelihood fit we obtain
$\scp = \finalscp$ and $\ccp = \finalccp$. 
 When we fix the values of 
$-\scp$ to the average \stwob measured in $\b\to\c\cbar\s$ modes,
$\stwob =  0.675 \pm 0.026$~\cite{hfag}, and $\ccp$ to zero,
and re-fit the data sample the negative log-likelihood changes by 2.2$\sigma$.
The signal yield is consistent with our findings in
the $B^0\to\KS\pi^+\pi^-$ decay~\cite{kelly} assuming isospin symmetry
and that the
dominant charmless final states are $f_0(980)\KS$,
$K^*(892)\piz$, $K^*_0(1430)\piz$, and non-resonant 
$\KS\piz\piz$.

We are grateful for the excellent luminosity and machine conditions
provided by our \pep2\ colleagues, 
and for the substantial dedicated effort from
the computing organizations that support \babar.
The collaborating institutions wish to thank 
SLAC for its support and kind hospitality. 
This work is supported by
DOE
and NSF (USA),
NSERC (Canada),
IHEP (China),
CEA and
CNRS-IN2P3
(France),
BMBF and DFG
(Germany),
INFN (Italy),
FOM (The Netherlands),
NFR (Norway),
MIST (Russia),
MEC (Spain), and
PPARC (United Kingdom). 
Individuals have received support from the
Marie Curie EIF (European Union) and
the A.~P.~Sloan Foundation.


\begin{thebibliography}{99}
\bibitem{cc} Unless explicitly stated, conjugate decay modes are
assumed throughout this paper.
\end{thebibliography}
\end{document}